\documentclass[twocolumn]{revtex4}
\usepackage[colorlinks,bookmarksopen,bookmarksnumbered,citecolor=red,urlcolor=blue]{hyperref}
\usepackage{moreverb}
\usepackage{graphicx}
\usepackage{epstopdf} 
\begin{document}


\title{Waxing and Waning of Observed Extreme Annual Tropical Rainfall}

\author{Jai Sukhatme and V. Venugopal}

\address{Centre for Atmospheric and Oceanic Sciences \& Divecha Centre for Climate Change, Indian Institute of Science, Bangalore 560012, India.}


\date{\today}

\begin{abstract}
We begin by  providing observational evidence that  the probability of
encountering  very high  and  very low  annual  tropical rainfall  has
increased  significantly  in  the   recent  decade  (1998-present)  as
compared to the preceding warming era (1979-1997).  These changes over
land and ocean are spatially  coherent and comprise of a rearrangement
of very  wet regions and a  systematic expansion of dry  zones.  While
the  increased likelihood  of  extremes is  consistent  with a  higher
average temperature during the pause (as compared to 1979-1997), it is
important to note  that the periods considered  are also characterized
by a transition from a relatively warm  to cold phase of the El Ni\~no
Southern Oscillation  (ENSO).  To  further probe the  relation between
contrasting phases  of ENSO  and extremes  in accumulation,  a similar
comparison is performed between 1960-1978 (another extended cold phase
of  ENSO)  and the  aforementioned  warming  era.  Though  limited  by
land-only observations, in this cold-to-warm transition, remarkably, a
near-exact  reversal  of  extremes  is noted  both  statistically  and
geographically.  This is despite  the average temperature being higher
in 1979-1997  as compared  to 1960-1978.   Taken together,  we propose
that there is a fundamental mode of natural variability, involving the
waxing  and waning  of  extremes in  accumulation  of global  tropical
rainfall with different phases of ENSO.
\vskip 0.2 truecm
\begin{center}
{\bf Journal Ref: QJRMS, DOI:10.1002/qj.2633, 2015.}
\end{center}

\end{abstract} 

\maketitle

\section{Introduction}
In the context  of global warming, the increasing  moisture content of
the troposphere  \citep{HS-2006,Trenberth-2011} is expected  to result
in  an   amplification  of  short-duration  extreme   rainfall  events
\citep{Trenberth-1999,AI-2002,GS-2009},  mostly validated  by regional
ground-based  observations   \citep{East-rev,Groisman-2005,Venu}.   On
longer timescales, a consequence of  the increase in column integrated
water vapour for locations with very  high and low accumulation is the
so-called thermodynamic  effect of  ``wet regions getting  wetter, and
dry   regions   getting  drier''   \citep{HS-2006,Wentz-2007}.    Some
observations  \citep{LP-2009,   Allan-etal-2010,  Chou-etal-2013}  and
long-term            global             warming            simulations
\citep{Chou-etal-2009,Giorgi-etal-2011,LWK-2013,Gorman-2012}       are
consistent  with  the  expected  consequences of  this  paradigm.   In
addition,  dynamical  changes  due  to warming  also  affect  rainfall
\citep{Seager-2010}.  The  combination  of these  two  complicate  the
precipitation  response  \citep{Xie-2010,Chad-2013},  especially  over
land \citep{Sonia2}.  In fact, as suggested by \cite{LA-2012,LA-2013},
the ``wet-wetter,  dry-drier" hypothesis may be  more appropriate over
land  if the  wet  and dry  regions  are not  considered  to be  fixed
geographical locations \citep[see also,][]{Polson-2013}.

Apart from the role of warming,  it has been suggested that changes in
regional extremes have a  natural component. In particular, individual
locations with more than a  century long data clearly exhibit multiple
cycles       in        heavy       rainfall       \citep[see       for
  example,][]{Willems-2013,Marani-2014}.    Further,   it   has   been
documented that regional extremes in  rainfall vary with El Ni\~no and
La Ni\~na conditions \citep[see][for reports on the continental United
  States,  South  America,  eastern  Australia  and  the  Philippines,
  respectively]{Gershunov-2003,Grimm-2009,  Alexander-2013, Phil}.  In
fact, connections between daily and monthly extremes and the El Ni\~no
Southern Oscillation (ENSO) on a  more global scale have been explored
over the tropical oceans \citep{Allan-Soden-2008} as well as over land
\citep{Lyon-2005,Curtis-2007,Alexander-2009}.       In
  addition to the effect of natural cycles on short-duration extremes,
  different regions  in the tropics experience anomalously  wet or dry
  years     during    El     Ni\~no    and     La     Ni\~na    events
  \citep{RH1,RH2,Dai-GRL}.

Thus, short-duration regional  rainfall extremes as well  as very high
and low annual  accumulation are plausibly influenced  by both warming
and natural cycles.  In the present work, we focus on the footprint of
ENSO  on annual  accumulation and  its  extremes in  the tropics.   In
Section  2, we  compare  extremes in  global  accumulation during  the
ongoing pause in global warming  and the preceding warming era.  Apart
from the  fact that the  average temperature during the  recent period
since   1998  is   higher  than   the  preceding   warming  era,   the
warming-vs-pause   contrast  is   also  a   comparison  between   long
predominantly  warm and  cold phases  of ENSO,  respectively.  Keeping
this  in  mind,  in  Section  3,  we  attempt  to  delineate  possible
connections between these changes in very low and high annual rainfall
and  ENSO phase  transitions.   We  also discuss,  in  Section 4,  the
consistency between  our global viewpoint  using annual rainfall  as a
measure,  and the  noted trends  in short-duration  regional extremes.
Finally, the  paper concludes with  a summary  of results and  a brief
discussion in Section 5.

\section{Warming vs Pause}

Observations  suggest   that,  despite   the  continual   build-up  of
greenhouse gases in the atmosphere,  the rate of surface warming since
1998  has  been slower  than  in  the preceding  decades  \citep{Fyfe,
  Cowtan2014}; a phenomenon  referred to as the  ``pause" or ``hiatus"
in  global warming.   While the  cause behind  the ongoing  hiatus has
received much attention, with many competing theories in the fray, the
answer remains  elusive \citep[see, for example,  the succinct summary
  in][]{Held-2013}. Rather  than worry about  its cause, here  we view
the pause  --- the first  of its kind  with possibly others  to follow
\citep{Meehl}  ---  as  a   natural  laboratory  wherein  the  climate
continues  to  evolve  with  one  its  primary  variables  being  held
relatively constant.   Given this unique  state of affairs,  the first
question we ask concerns the fate of tropical rain during the pause as
compared to the immediately  preceding warming era \citep[beginning in
  the late 1970s,][]{TF-2013}.

With  regard  to   rainfall  measurements,  the  Global  Precipitation
Climatology  Product (GPCP) provides  data at  a spatial  and temporal
resolution    of    2.5   degrees    and    1   month,    respectively
\citep{adler-etal-2003}.  This  data is  available  from  1979 to  the
present day,  thus covering the  hiatus and the preceding  warming era.
Due  to its monthly
temporal  and  coarse  spatial  resolution,  short-duration  localized
intense precipitation events which are usually the focus of studies on
extremes lie outside the scope of this data. Rather, the measure which
we focus  on in this  work, which is  arguably a better  yardstick for
assessing   the   ``wet-wetter,   dry-drier''   paradigm,   is   annual
accumulation  at every  grid point.  Thus,  in the  remainder of  this
manuscript, extremes refer to very high and low annual accumulation.

\begin{figure}
\centering
\includegraphics[width = 0.4\textwidth]{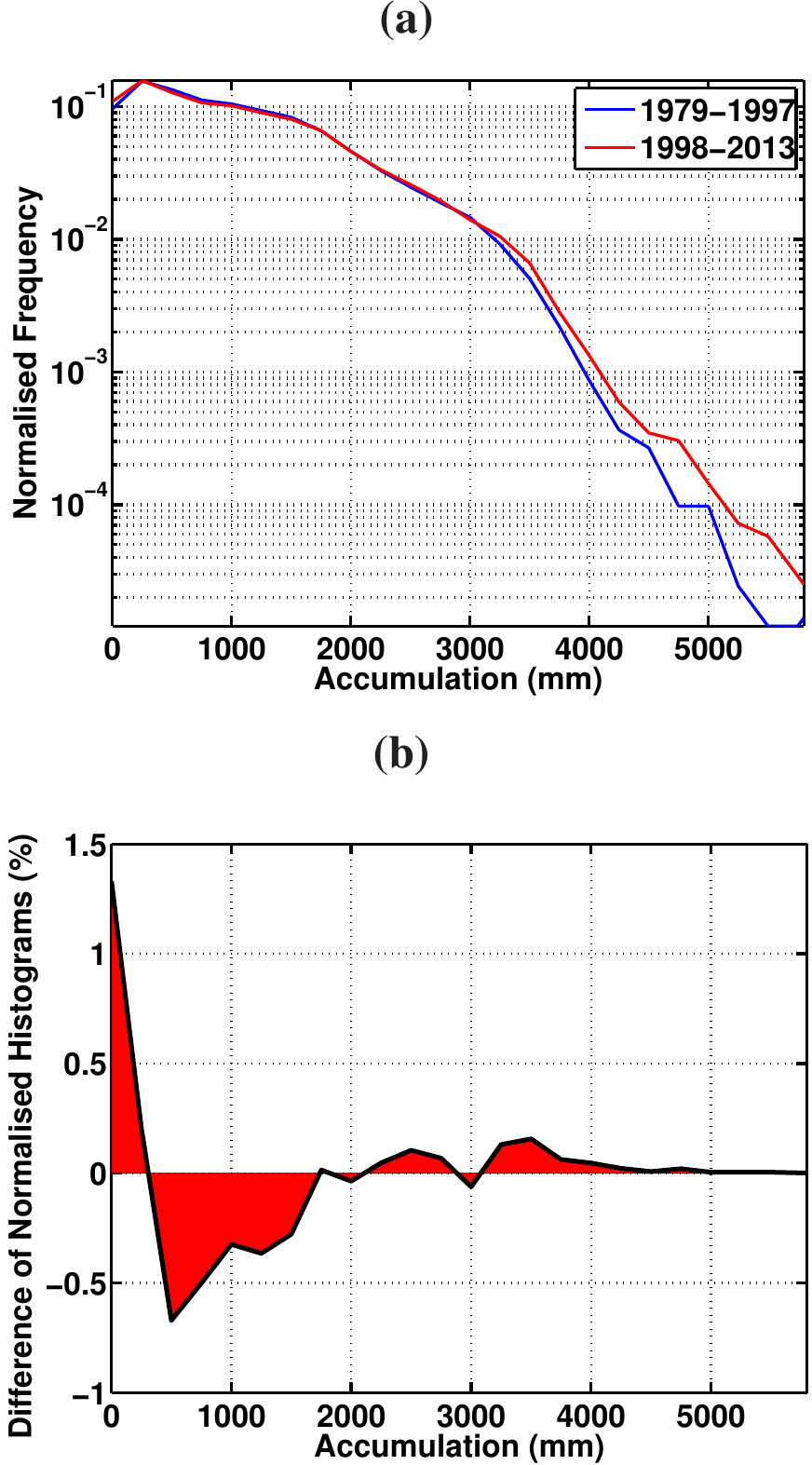}
\caption{(a)  Normalised  frequency  histograms  of  the
  annual  accumulation  (mm)  during  1979-1997 (blue)  and  1998-2013
  (red);  and (b)  Differences of  these histograms.   The frequencies
  shown are based on the union of  the data from each of the two eras,
  i.e., sample sizes  of 19 (16 years) $\times$  28 (35S-35N) $\times$
  144  (0-360  longitude).   The   annual  accumulation  is  based  on
  2.5-degree, monthly GPCP rainfall.}
\label{fig:fig1}  
\end{figure}

\begin{figure*}
\centering
\includegraphics[width = 0.8\textwidth]{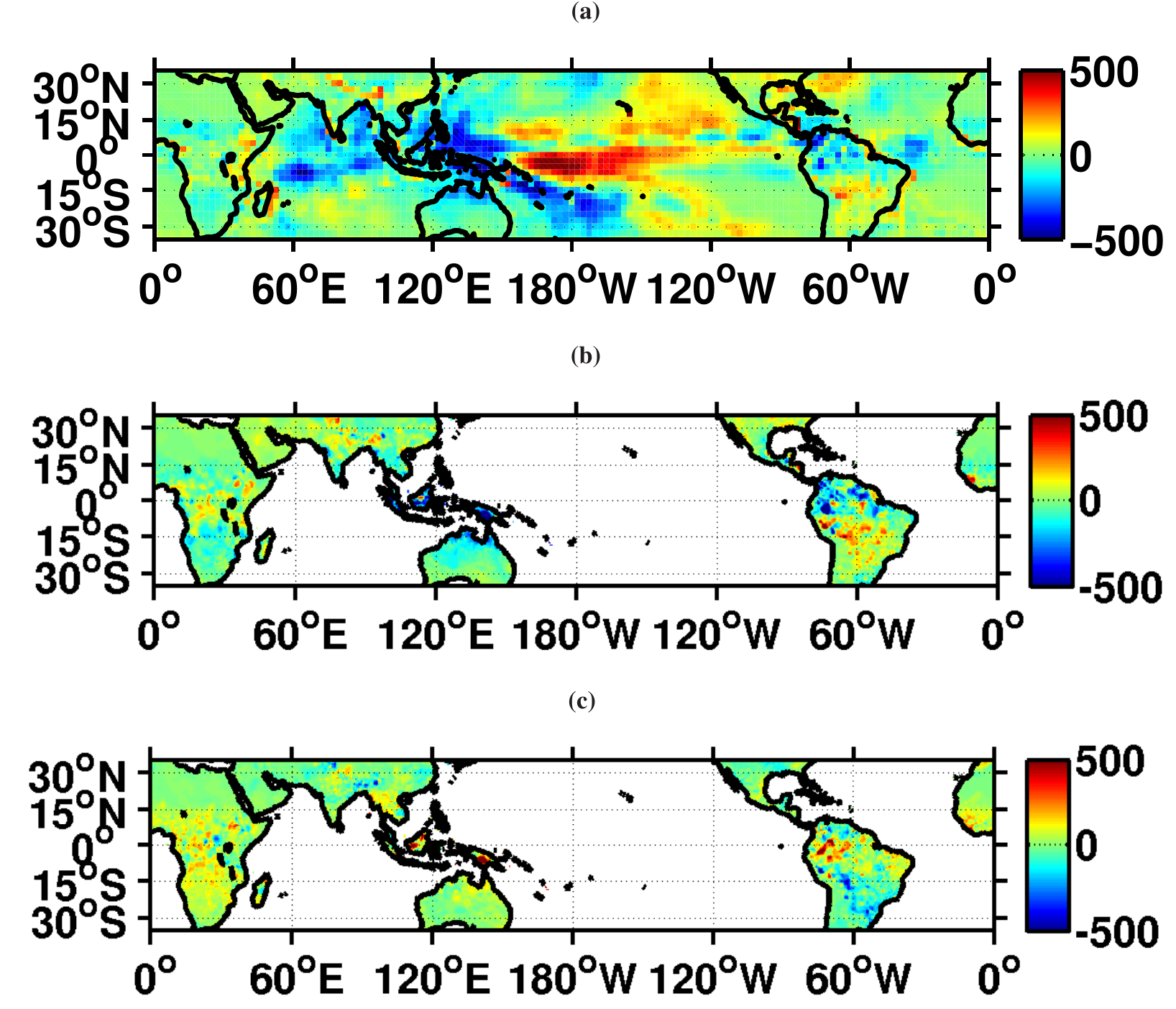}
\caption{(a) Spatial distribution  of the  difference of
  climatologies of  annual accumulation (Warming - Pause; in mm),  based on
  monthly, 2.5-degree, GPCP rainfall. (b) Same as (a), but over land
  using 0.5-degree, GPCC rainfall. (c) Same as (b), but between
  1960-1978 and 1979-1997 (former - latter period).}
\label{fig:fig2}  
\end{figure*}

We  begin by examining  the differences  in annual  tropical (35S-35N)
rainfall accumulation as recorded in the hiatus (1998 to 2013) and the
preceding warming  era (1979  to 1997).  Figure  \ref{fig:fig1}a shows
the normalised frequency distributions of annual rainfall during these
two periods.  These  distributions are based on the  union of the data
from every year of the respective  era, i.e., a sample size of 19 (16)
$\times$ 28  $\times$ 72.  While frequency distributions  are used for
illustrating changes, cumulative distribution functions (CDF) are used
for significance  testing.  Specifically, the  Kolmogorov-Smirnov (KS)
test \citep{Papoulis} shows that the CDFs of accumulation have changed
significantly between the two eras  (a KS distance of 0.02, leading to
a  $p$-value  close to  zero,  based  on  the null  hypothesis  $H_0$:
CDF$_{\rm  {hiatus}}$ =  CDF$_{\rm {warming}}$).   The changes  in the
tails   of   the  distributions   are   better   captured  in   Figure
\ref{fig:fig1}b, which shows their  difference (hiatus - warming).  In
particular, we note that, in  the global tropics, the probabilities of
encountering  very  low ($<$  200  mm) and  very  high  ($>$ 3000  mm)
accumulation have  increased significantly  during the hiatus.   It is
worth reiterating that  in this comparison we are  considering all the
years  that make up  an era  to be  a single  set, thus  the increased
probability of  extremes in  accumulation is true  of the hiatus  as a
whole    and    individual     years    can    deviate    from    this
expectation.  


In addition to the pause being warmer (on average) than the preceding
warming  era,  the periods  considered  are  also  characterized by  a
transition  from  a  relatively  warm   to  cold  phase  of  the  ENSO
\citep{TF-2013}.   This  transition is  evident  when  we examine  the
difference  in rainfall  climatology between  the two  eras  (shown in
Figure  \ref{fig:fig2}a).  For  example,  we see  an east-west  anomaly
along the  equatorial Pacific ocean, indicating a  preference of moist
convection  more  to the  west  (east)  during  the hiatus  (warming).
Similarly, the  subtropical signature  of the two  phases can  be seen
with the southeastern spreading of anomalies into the southern Pacific
Ocean \citep[see, for example,][]{RH1,RH2,Wallace-ENSO,Dai-GRL}.

Thus, as both warming  and a phase transition in ENSO  are in play for
the  present comparison,  it is  not possible  to attribute  the noted
increase  in extremes  of tropical  accumulation to  any one  of these
factors  \citep[see also  the  discussion in][for  possible models  of
  changes in rainfall distribution due to ENSO and warming]{Pend}.  To
focus on the role of phase changes in ENSO on extreme accumulation, we
note that  the era  preceding the  late seventies,  i.e., 1960  to the
mid-1970s, was also  a long cold phase of ENSO  \citep[see for example
  the discussion in][]{ZWB}.  In fact, both these transitions are also
captured by the  Pacific Decadal Oscillation index  (see, for example,
\url{http://jisao.washington.edu/pdo}).   It should  be  kept in  mind
that a period identified as a  particular phase of ENSO is interrupted
by events of opposite polarity.  For example, even though the pause is
by and  large a cold  phase of ENSO, an  examination of the  PDO index
reveals  that  the  hiatus  too  can  be  partitioned  into  pre-  and
post-2005,  where  the  latter  period   is  dominated  by  La  Ni\~na
conditions.

Given that there are no long-term tropical rainfall observations which
cover both land and ocean, we utilise Global Precipitation Climatology
Centre (GPCC) data \citep{gpcc}, that is based on station observations
(only  land) and  is available  from 1950  onwards, at  a spatial  and
temporal resolution of $0.5^\circ$ degree and 1 month, respectively. A
difference in climatologies  of annual accumulation for  the two phase
transitions   of  ENSO,   namely,   (i)   \{1979-to-1997\}  {\it   vs}
\{1998-to-2013\}  (warm-to-cold) and  (ii)  \{1960-to-1978\} {\it  vs}
\{1979-to-1997\}  (cold-to-warm) are  shown in  Figure \ref{fig:fig2}b
and  c,  respectively.  Not  only  are  changes over  land  consistent
between  Figures \ref{fig:fig2}a  and  b, they  are  also opposite  in
character to those shown in Figure \ref{fig:fig2}c.

Having  noted  the  expected  changes in  climatologies,  the  specific
question we  seek to  answer is whether  a cold-to-warm  transition is
characterised  by a  decrease in  extremes; i.e.,  is there  a natural
modulation  of very  high and  low accumulation  associated  with ENSO
phase transitions?   In other words,  given that the  mean temperature
during  1979-1997 was  higher than  the  preceding era  (1960s to  the
mid-1970s),  if there  is  indeed  a decrease  in  global extremes  of
accumulation, it clearly points to the role of ENSO phase transitions.

\section{Extremes and ENSO Transitions}

\begin{figure}
\centering
\includegraphics[width = 0.4\textwidth]{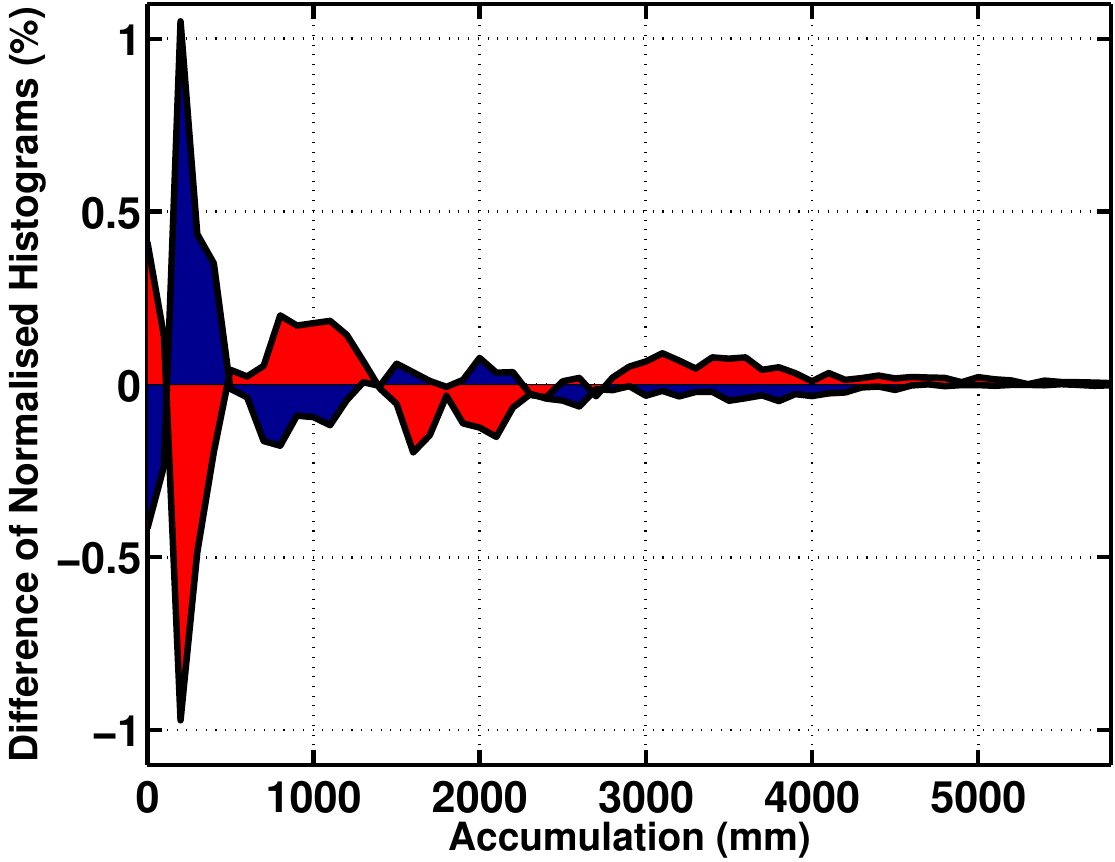}
\caption{Differences  in  the  normalised  histograms  of  the  annual
  accumulation   over   land   for   two   ENSO   phase   transitions,
  \{1979-to-1997\}$-$\{1960-to-1978\}  (blue fill;  cold to  warm) and
  \{1998-to-2013\}$-$\{1979-to-1997\}  (red fill;  warm to  cold). The
  estimates are  based on  GPCC 0.5-degree, monthly  observations over
  land.}
\label{fig:fig3}
\end{figure}

Figure  \ref{fig:fig3} shows  the differences  between the  normalised
histograms  of  accumulation,  based  on  GPCC  data,  for  two  phase
transitions  of ENSO:  (i) \{1979-to-1997\}  $-$ \{1960-to-1978\}  and
(ii) \{1998-to-2013\} $-$ \{1979-to-1997\}.   We note that the changes
in very high and low accumulation for the warm-to-cold transition from
both  GPCP  (Figure \ref{fig:fig1}b)  and  GPCC  (red fill  in  Figure
\ref{fig:fig3}) are  qualitatively similar. Further, even  though land
and  ocean rainfall  estimates from  GPCP have  different biases,  the
consistency  with GPCC  is reassuring.   It is  worth noting  that the
percentage changes in Figure \ref{fig:fig3} are smaller than in Figure
\ref{fig:fig1}b.   This  could be  attributed  to  the higher  spatial
resolution  of GPCC  observations.   More strikingly,  the changes  in
extremes from  a cold \{1960-to-1978\} to  warm \{1979-to-1997\} phase
of ENSO (blue fill in  Figure \ref{fig:fig3}) are exactly the opposite
of what  is observed  in a warm-to-cold  transition.  This  shows that
relative  changes in  extremes  of tropical  accumulation are  closely
linked to ENSO.

To  ascertain if  there is  a  spatial character  associated with  the
statistical  changes  described above,  we  utilize  the crossings  at
approximately  200  mm  and  3000  mm in  Figures  \ref{fig:fig1}  and
\ref{fig:fig3} as thresholds for very low (``dry'') and high (``wet'')
accumulation,  respectively.  Further,  the two  eras that  straddle a
particular transition are denoted by E1 and E2 (where E2 follows E1 in
time). Using this terminology, we construct ``Index maps" that consist
of  a  union  of  three  sets. Specifically,  these  maps  consist  of
geographical locations that accumulated  more (less) than 3000 mm (200
mm) of rain in (i) one or more  of the years of E1 and E2 (cyan); (ii)
one or  more of the years  of E2 and none  of E1 (blue);  (iii) one or
more of  the years of  E1 and  none of the  years in E2  (red).  Thus,
locations in  blue (red)  represent appearance (disappearance)  of wet
and dry regions in E2 when compared to E1.

\begin{figure*}
\centering
\includegraphics[width = 0.8\textwidth]{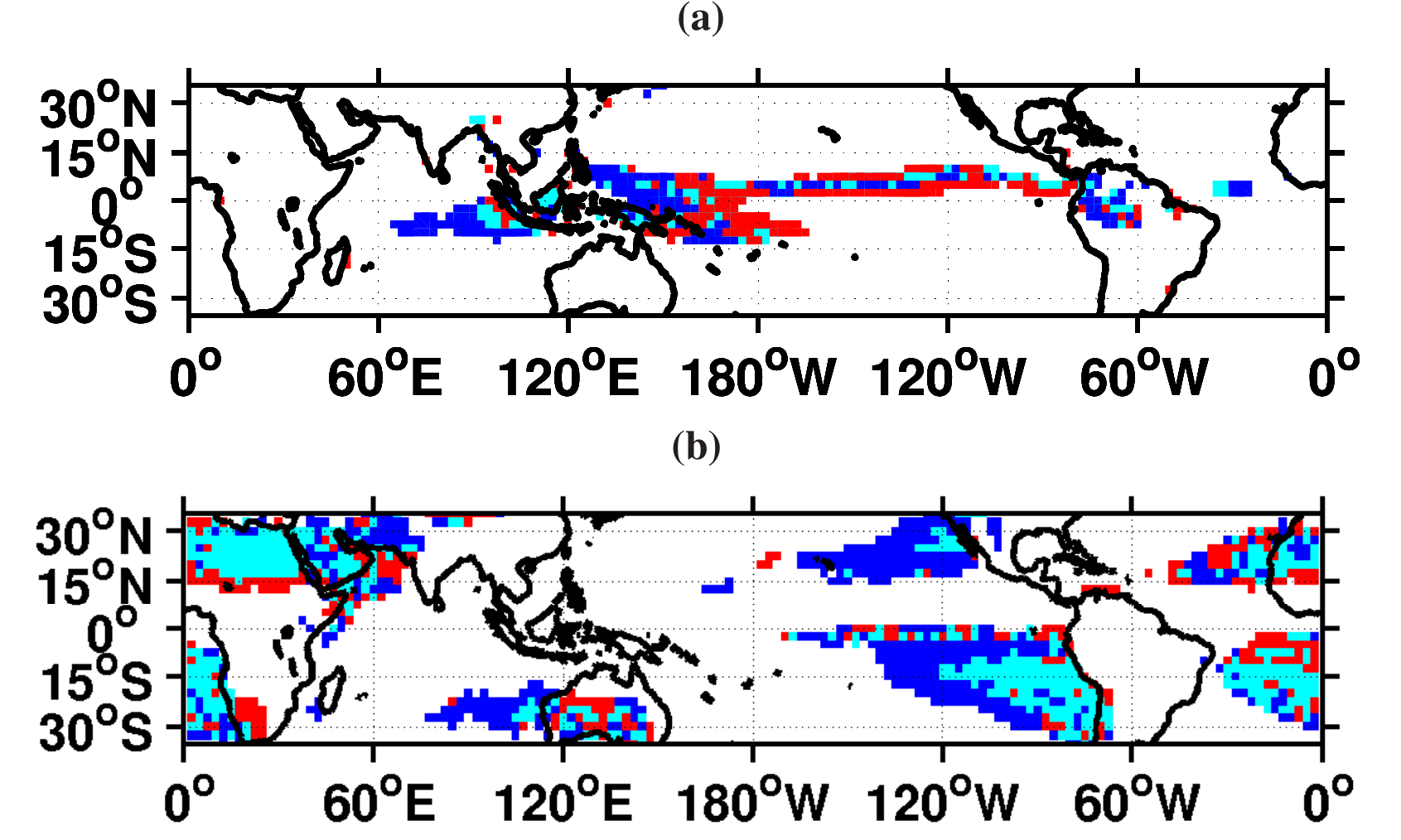}
\caption{Spatial distribution of the changes in the climatological (a)
  wet and  (b) dry regions in  the tropics, based  on GPCP 2.5-degree,
  monthly   observations.    The   colours   shown   represent   those
  geographical  locations that  accumulated more  (less) than  3000 mm
  (200 mm) of rain (i) in one or more of the years of \{1998-to-2013\}
  and  none of \{1979-to-1997\}  (blue); (ii)  in one  or more  of the
  years of \{1979-to-1997\} and  none of the years in \{1998-to-2013\}
  (red); and (iii) in one or more of the years of both eras (cyan).}
\label{fig:fig4}
\end{figure*}

\begin{figure*}
\centering
\includegraphics[width=0.75\textwidth]{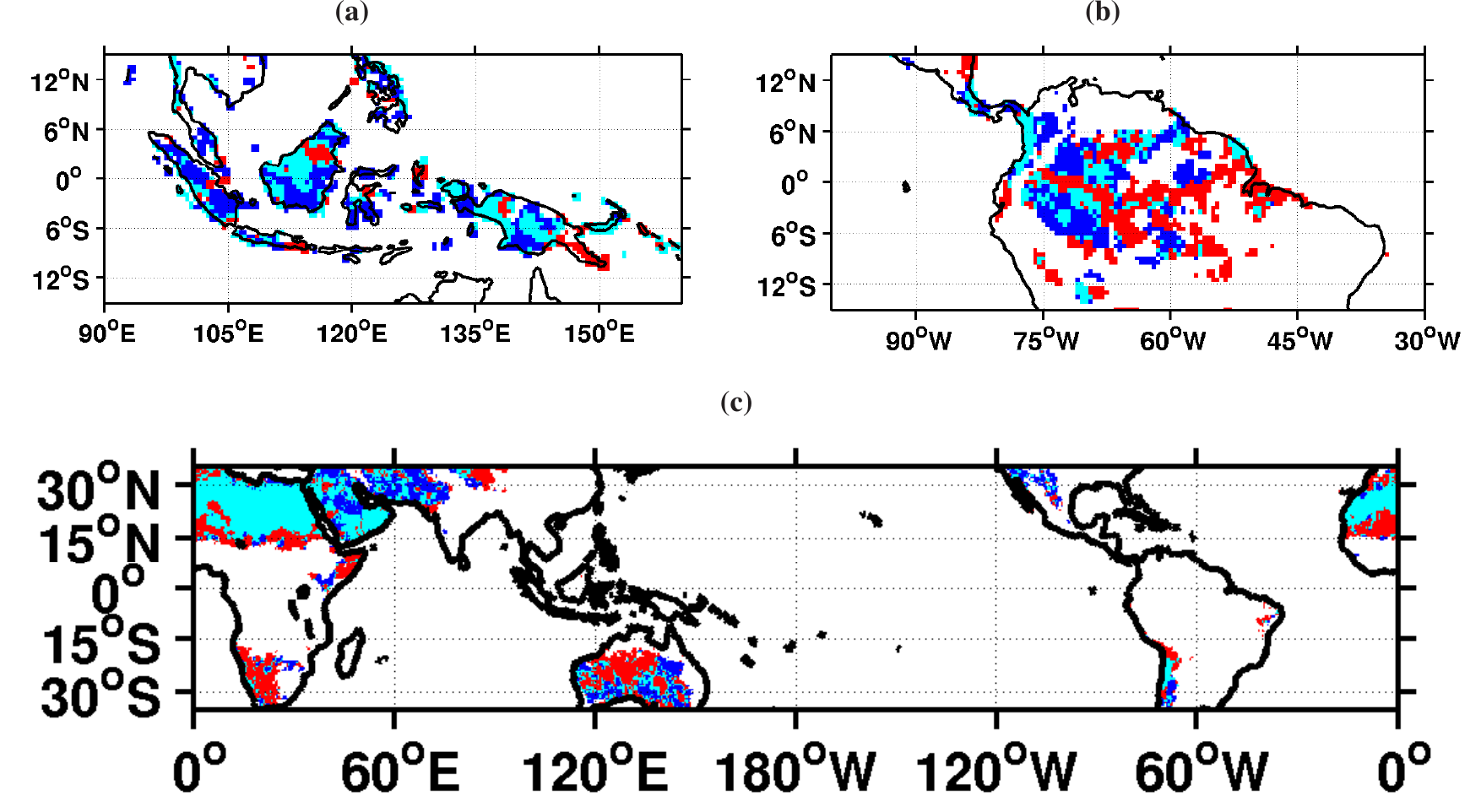}
\caption{Construction and color  scheme same as Figure \ref{fig:fig4},
  but based on GPCC 0.5-degree, monthly land-only observations for the
  eras \{1979-to-1997\} and \{1998-to-2013\}.  (a, b) Wet regions; (c)
  Dry regions. As before, blue/red represents appearance/disappearance
  of new wet/dry locations, and cyan represents no change.}
\label{fig:fig5}
\end{figure*}

Consider first, the warm-to-cold transition, i.e., E1=\{1979-to-1997\}
and  E2=\{1998-to-2013\}, which  is  captured by  both  GPCP and  GPCC
data. As  seen from the  global GPCP product  (Figure \ref{fig:fig4}),
new  high  accumulation  regions  appear  near  the  western  maritime
continent, over  the Indian Ocean  and the western part  of equatorial
South America  (blue in Figure \ref{fig:fig4}a).   This is accompanied
by a  depletion in  the eastern core  of the Pacific  convergence zone
(red).    Overall,   the    ``new''   locations   still   lie   within
climatologically rainy  zones, thus indicating a  rearrangement of wet
regions.  The aforementioned changes over land are also seen from GPCC
(blue in  Figures \ref{fig:fig5}a,b).  At  the other end, the  new dry
points  take the  form of  spatially coherent  fringes and  indicate a
systematic  expansion of  the  existing dry  regions  off the  western
coasts  of Australia,  and North  and  South America  (blue in  Figure
\ref{fig:fig4}b).   As  this warm-to-cold  transition
  also  involves  comparison  with on-an-average  warmer  temperatures
  (during the  pause), it  is worth noting  that the expansion  of dry
  zones     is     sometimes      linked     to     global     warming
  \citep{Marvel}\footnote{This   expansion   of  dry
    zones is  part of a developing  story on the  widening of tropical
    Hadley  cells, which  it  itself  appears to  have  a strong  ENSO
    component \citep[see, for example,][]{Joy-new}.}. Furthermore, on
land (most clearly seen in Figure \ref{fig:fig5}c from GPCC due to its
higher  spatial  resolution),  we  observe  a mixed  signal  over  the
Australian continent  (red and  blue), signs of  new dry zones  in the
western  United States and  northern Pakistan  (blue), along  with the
disappearance  of  existing  dry  regions  in southern  Africa  and  a
coherent  shrinking of  the deserts  in the  Sahel region  of northern
Africa (red).  It is worth noting  that these changes over land are in
very good agreement with regional reports; specifically, the projected
drying   of   the   western   United   States   \citep[2000   onwards;
  see][]{Seager-etal-2007}, the occurrence of repeated severe droughts
over northern Pakistan in the early 2000s\footnote{See for example the
  documentation        of        droughts        over        Pakistan,
  \url{http://pakistanweatherportal.com/2011/05/08/
    history-of-drought-in-pakistan-in-detail}.}, and  the reduction of
dryness  over  southern  Africa  \citep[1998  onwards;  see  Figure  2
  in][]{Giannini-2008} as well as the Sahel \citep{sahel}.

\begin{figure*}
\centering
\includegraphics[width = 0.75\textwidth]{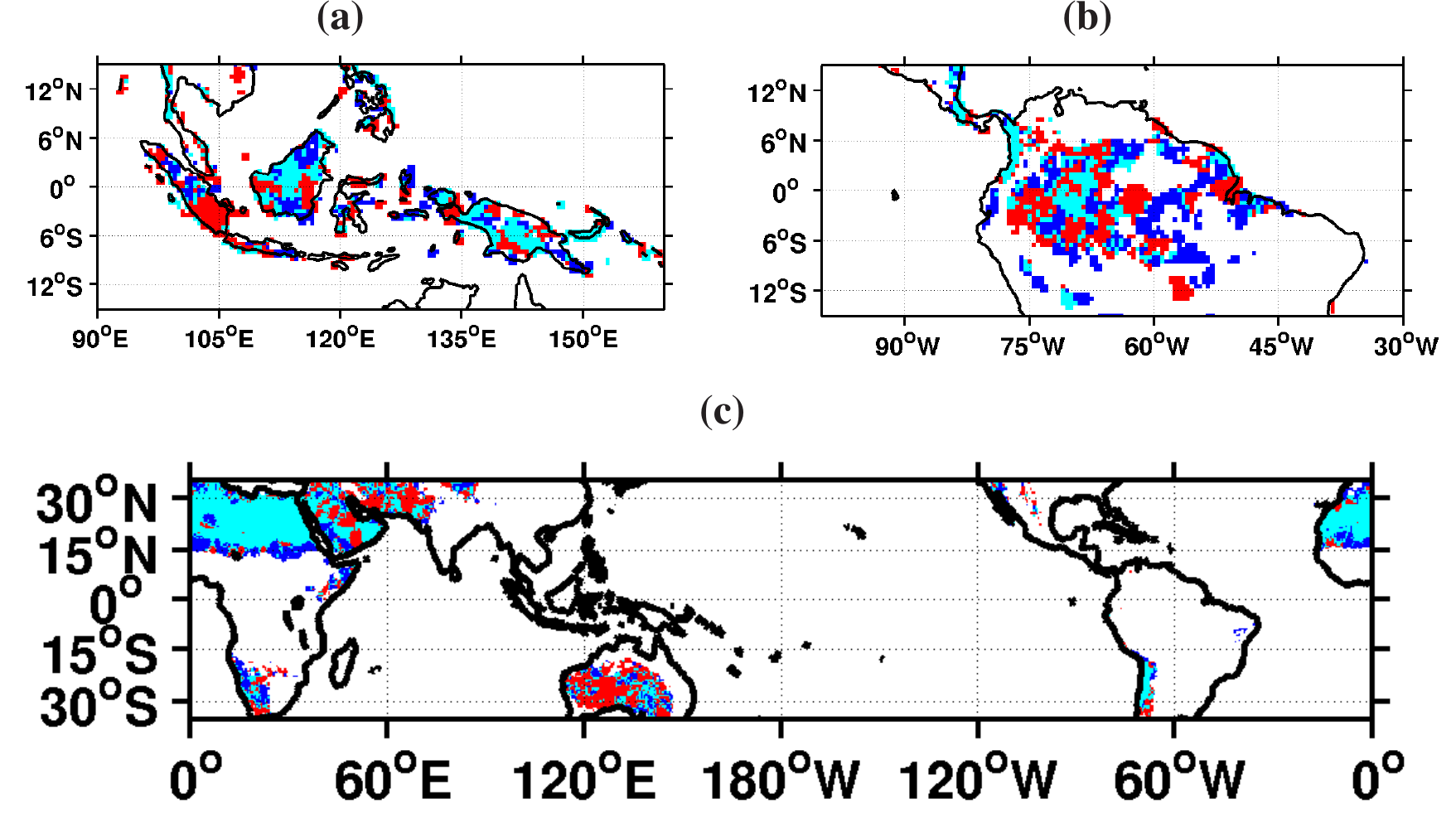}
\caption{Same as Figure \ref{fig:fig5}, but for the
  eras \{1960-to-1978\} and \{1979-to-1997\}.  (a, b) Wet regions; (c)
  Dry regions. As before, blue/red represents appearance/disappearance
  of new wet/dry locations, and cyan represents no change.}
\label{fig:fig6} 
\end{figure*}

In   the  cold-to-warm   transition  (i.e.,   E1=\{1960-to-1978\}  and
E2=\{1979-to-1997\}),  which  is  only  captured  by  GPCC  data,  the
geographical  changes (over  land) seen  in Figure  \ref{fig:fig6} are
almost   exactly   opposite   in   character  to   those   in   Figure
\ref{fig:fig5}.  These changes are  consistent with the reversal seen
in  the   two  ends  of   the  probability  distributions   in  Figure
\ref{fig:fig3}.   In  particular,   the  maritime  continent  shows  a
disappearance of  wet points  to the west  (red), as does  the western
portion of equatorial South  America (red). The same observation holds
true for the very dry zones; here, new dry points are seen in southern
Africa  and  the  Sahel  region  (blue), while  dryness  reduces  over
Pakistan,  western United  States  and much  of continental  Australia
(red).

Thus, the answer to the question  posed earlier (at the end of Section
2) is  that, there appears  to be an  intrinsic waxing and  waning, of
very  high  and  low  tropical  rainfall associated  with  ENSO  phase
transitions. Not only are  the changes statistically significant, they
also bear a coherent spatial signature. 

\section{Consistency with trends in regional short-duration extremes}

Having taken a global point of view, with a focus on
  annual accumulation, we now assess whether the changes seen above
  agree with previously reported trends in regional
  extremes.  Beginning with \cite{IY}, there have been numerous
reports of increasing trends in short-duration extreme rainfall events
across the globe \citep[see, for
  example,][]{East-rev,Groisman-2005,Venu}.  In order to show that
this increase is in fact consistent with our global accumulation
picture (waxing and waning), we use North America (15N-45N; 60W-130W)
as an example.  The motivation behind this choice stems from the fact
that the region considered is large enough, so as to smooth out large
local fluctuations, and thus more amenable to study extremes
\citep[see, for example,][and the references therein]{Groisman-2005}.

\begin{figure*}
\centering
\includegraphics[width=0.4\textwidth]{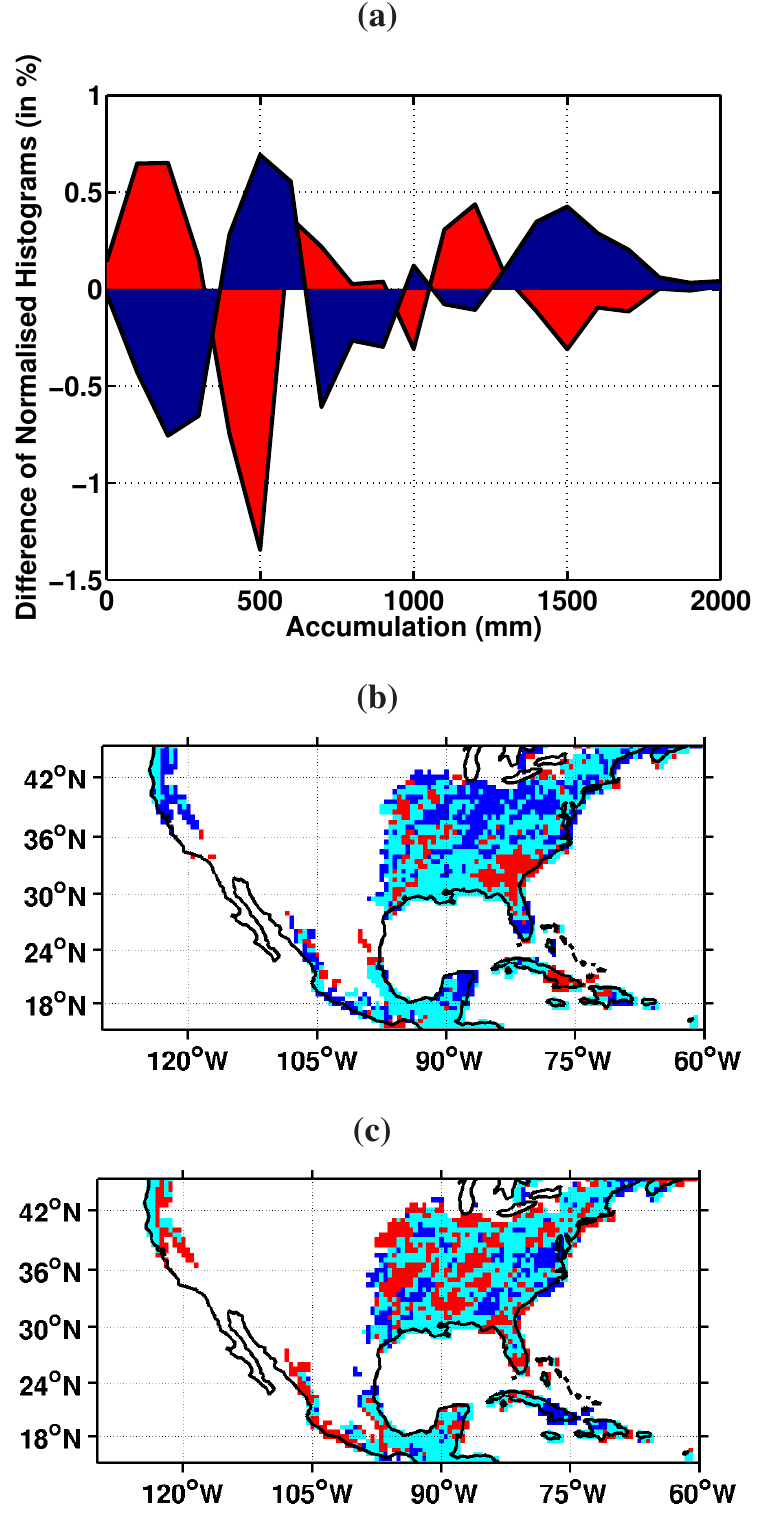}
\caption{(a)  Same as  Figure  \ref{fig:fig3}, but  for North  America
  (blue fill: cold-to-warm;  red fill: warm-to-cold).  Construction of
  the index  map in the  lower two panels  is based on a  threshold of
  1200  mm or  above, using  GPCC land-only  observations.   The color
  scheme   is   the   same   as   in   Figure   \ref{fig:fig6}.    (b)
  \{1960-to-1978\}  and  \{1979-to-1997\};  (c)  \{1979-to-1997\}  and
  \{1998-to-2013\}.   As  before,  in  panels (b)  and  (c),  blue/red
  represents appearance/disappearance of new wet/dry locations, and
  cyan represents no change.}
\label{fig:fig7}
\end{figure*}

Following the  same procedure as  before, we construct  histograms and
index  maps for  the  two  phase transitions  over  this region.   The
difference  seen in  the  right  tails of  the  histograms during  the
cold-to-warm   transition  (blue   fill  in   Figure  \ref{fig:fig7}a)
indicates  an increased  likelihood  of exceeding  annual rainfall  of
$\sim$1200 mm.   Indeed, this increase  is marked by an  appearance of
new wet spots of high accumulation (blue in Figure \ref{fig:fig7}b) in
\{1979-to-1997\}  as  compared to  \{1960-to-1978\}  over central  US.
This  matches precisely  with the  significant  short-duration extreme
event trends from the mid 1970s to the late 1990s, over the contiguous
central US, reported by  \cite{Groisman-2005}.  Furthermore, as in the
case  of  global  tropics,  the  statistical  and  geographic  changes
associated with a warm-to-cold transition (Figure \ref{fig:fig7}c) are
opposite in nature to those during a cold-to-warm transition.

At  first  glance,  in   the  cold-to-warm  transition,  the  increase
(decrease)  in  likelihood  of  encountering very  high  (low)  annual
rainfall over North America might  appear to be contrary to the global
tropical perspective  presented earlier (i.e., waning of  both wet and
dry  accumulation extremes;  blue fill  in Figure  \ref{fig:fig3}).  A
closer examination, however, indicates that  this is not the case.  In
fact, the very high accumulation  in the North American region forms a
subset of the middle portion  of the global histogram, and the changes
between $\sim$1200 mm and $\sim$2000 mm in Figures \ref{fig:fig7}a and
\ref{fig:fig3} are similar.

\section{Discussion}
An immediate implication  of our finding is that a  warming signal can
be  enhanced or  subdued depending  on  the phase  of ENSO.   However,
disentangling their  respective contributions  to changes  in rainfall
extremes remains  a challenge \citep[e.g.,][]{Shukla}.  That  said, it
is worth  asking if our  methodology can also  shed light on  the more
gradual contribution of warming. To this end, a comparison of extremes
in similar phases  of ENSO could prove  fruitful.  Specifically, given
the data at  hand, we examined the changes in  extreme accumulation in
two cases:  (i) between two  temporally well-separated cold  phases of
ENSO (1960-1978  vs 1998-2013); and  (ii) within the warming  era that
was mostly  a warm phase  of ENSO  (1979-1987 vs 1988-1997).   In both
these  cases, the  average  temperature is  higher  in the  respective
latter period.  The first experiment  (whose results can be deduced by
summing the two curves in Figure  3) yields a marginal increase on the
very wet  side, and, in fact,  a decrease in dry  extremes, neither of
which is statistically significant. The second experiment (differences
in PDFs not shown) showed almost no change in extremes on either side;
in fact,  if anything, both  very high  and low accumulation  showed a
marginal decrease.   Taken together,  it is difficult  to argue  for a
clear influence of warming on  the extremes of annual accumulation; in
fact, internal variability, as governed by the phases of ENSO, appears
to play a much more significant role than warming.

To summarise,  we studied the  changes in  very low and  high tropical
rainfall accumulation from a global point of view. The main finding is
that there appears to be a  fundamental natural mode of variability in
tropical rainfall  accumulation extremes  with the changing  phases of
ENSO.   Specifically,   our  analysis  provides   clear  observational
evidence  that a  warm-to-cold  (cold-to-warm) transition  of ENSO  is
associated  with   waxing  (waning)  of  extreme   accumulation,  both
statistically and  spatially, over  the global tropics.   The dominant
role  of ENSO  was made  clear  by comparing  accumulation across  two
transitions; specifically, cold-to-warm (1960-1978 vs 1979-1997) and a
warm-to-cold  (1979-1997  to  1998-2013),  both of  which  involved  a
progressive increase in  average global temperatures\footnote{In fact,
  recent   work  compares   accumulation  within   the  pause   (where
  temperatures are  fairly uniform),  and once  again, an  increase in
  extremes  due  to  a  warm-to-cold  transition  is  clearly  evident
  \citep{VS}.}.   Moreover, as  illustrated over  the continental  US,
this global  modulation is consistent with  previously reported trends
in short-duration regional extremes.

\vskip 0.5truecm {\it Acknowledgements:} JS \& VV acknowledge financial support from the
Divecha Centre  for Climate  Change.  Discussions with  George Huffman
and  John M.   Wallace are  greatly  appreciated. We  thank the  Earth
System Research  Laboratory, NOAA for making  available long-term GPCP
and  GPCC  precipitation observations.   We  thank  the two  anonymous
reviewers for their insightful and constructive comments.

\end{document}